# LOSS OF LANDAU DAMPING FOR BUNCH OSCILLATIONS

A. Burov, FNAL*, Batavia, IL 60510, U.S.A.

*Abstract*

Conditions for the existence, uniqueness and stability of self-consistent bunch steady states are considered. For the existence and uniqueness problems, simple algebraic criteria are derived for both the action and Hamiltonian domain distributions. For the stability problem, van Kampen theory is used [1-3]. The onset of a discrete van Kampen mode means the emergence of a coherent mode without any Landau damping; thus, even a tiny couple-bunch or multi-turn wake is sufficient to drive the instability. The method presented here assumes an arbitrary impedance, RF shape, and beam distribution function. Available areas on the intensity-emittance plane are shown for resistive wall wake and single harmonic, bunch shortening and bunch lengthening RF configurations. Thresholds calculated for the Tevatron parameters and impedance model are in agreement with the observations. These thresholds are found to be extremely sensitive to the small-argument behaviour of the bunch distribution function. Accordingly, a method to increase the LLD threshold is suggested. This article summarizes and extends recent author's publications [4, 5].

## MAIN EQUATIONS

Let $H(z,p)$ be a Hamiltonian for longitudinal motion inside an RF bucket distorted by the wake field:

$$H(z,p)=\frac{p^2}{2}+U(z)+V(z,t)\,;$$
$$U(z)=U_{\rm rf}(z)-\int\lambda(z')W(z-z')dz'\,; \qquad (1)$$
$$V(z,t)=-\int\rho(z',t)W(z-z')dz'\,.$$

Here $z$ and $p$ are the offset and the momentum of a particle, $U(z)$ is the steady state potential with $U_{\rm rf}(z)$ as its RF part, $\lambda(z)$ is steady state linear density, $W(z)$ is the wake function, $V(z,t)$ and $\rho(z,t)$ are small perturbations of the potential and linear density. For the potential well $U(z)$, action $I$ and phase $\varphi$ variables can be found:

$$I(H)=\frac{1}{\pi}\int_{z_{\min}}^{z_{\max}}\sqrt{2(H-U(z))}dz\,;$$
$$\Omega(I)=\frac{dH}{dI};\quad \frac{dz}{d\varphi}=\frac{\sqrt{2(H-U(z))}}{\Omega(I)}\,. \qquad (2)$$

The linear density $\lambda$ and its perturbation $\rho$ can be related to steady state phase space density $F(I)$ and its perturbation $f(I,\varphi,t)$:

$$\lambda(z)=\int F(I)dp\,;$$
$$\rho(z,t)=\int f(I,\varphi,t)dp\,. \qquad (3)$$

Below, the steady state distribution $F(I)$ is treated as an input function, determined either by cooling-diffusion kinetics, or by injection. The perturbation $f(I,\varphi,t)$ satisfies the Boltzmann-Jeans-Vlasov (BJV) equation [6]:

$$\frac{\partial f}{\partial t}+\Omega(I)\frac{\partial f}{\partial\varphi}-\frac{\partial V}{\partial\varphi}F'(I)=0\,. \qquad (4)$$

Equations (1-4) assume given input functions $U_{\rm rf}(z)$, $W(z)$ and $F(I)$, with $F'(I)\equiv dF/dI$, while the steady state solution $U(z)$, $I(H)$, $\lambda(z)$ and all the eigenfunctions of the BJV equation (4) are to be found.

To obtain the steady state solution, the following set of three equations must be solved:

$$U(z)=U_{\rm rf}(z)-\int\lambda(z')W(z-z')dz'\equiv U_{RHS}[\lambda];$$
$$I(H)=\frac{1}{\pi}\int_{z_{\min}}^{z_{\max}}\sqrt{2(H-U(z))}dz\equiv I_{RHS}[U]; \qquad (5)$$
$$\lambda(z)=2\int_{U(z)}^{H_{\max}}\frac{F(I(H))}{\sqrt{2(H-U(z))}}dH\equiv\lambda_{RHS}[I,U].$$

For any given input functions $U_{\rm rf}(z)$, $W(z)$ and $F(I)$, the solution can be obtained numerically by means of the relaxation method. Indeed, let it be assumed that initially there is no wake, so that the entire potential well is equal to the RF potential $U(z)=U_0(z)=U_{\rm rf}(z)$. With that assumption, initial action and linear density functions $I_0(H)$ and $\lambda_0(z)$ can be found from the 2nd and 3rd equations of the set (5). Then the following iteration procedure can be applied:

$$U_n(z)=U_{n-1}(z)-\varepsilon\left(U_{n-1}(z)-U_{RHS}[\lambda_{n-1}]\right);$$
$$I_n(H)=I_{RHS}[U_n]; \qquad (6)$$
$$\lambda_n(z)=\lambda_{RHS}[I_n,U_n]\quad;\quad n=1,2,....$$

If the solution exists and the convergence parameter $\varepsilon>0$ is sufficiently small, the process is very likely to converge. . The stability of the steady state solution can be determined by analysis of the BJV equation (4). Following Oide and Yokoya [7], the eigenfunctions may be expanded in Fourier series over the synchrotron phase $\varphi$:

$$f(I,\varphi,t)=e^{-i\omega t}\sum_{m=1}^{\infty}\left[f_m(I)\cos m\varphi+g_m(I)\sin m\varphi\right]. \quad (7)$$

With the zero-phase at the left stopping point,

$$\begin{aligned}&z(I,\varphi=0)=z_{\min}(I);\ z(I,\varphi=\pi)=z_{\max}(I);\\&z(I,-\varphi)=z(I,\varphi);\ p(I,-\varphi)=-p(I,\varphi),\end{aligned} \quad (8)$$

yield an equation for the amplitudes $f_m(I)$:

$$\begin{aligned}&\left[\omega^2-m^2\Omega^2(I)\right]f_m(I)=\\&-2m^2\Omega(I)F'(I)\sum_{n=1}^{\infty}\int dI'V_{mn}(I,I')f_n(I');\\&V_{mn}(I,I')=\\&-\frac{2}{\pi}\int_0^{\pi}d\varphi\int_0^{\pi}d\varphi'\cos(m\varphi)\cos(n\varphi')W(z(I,\varphi)-z(I',\varphi')).\end{aligned} \quad (9)$$

The matrix elements $V_{mn}(I,I')$ can be also expressed in terms of the impedance $Z(q)$. After [8, Eq. (2.69)]

$$W(z)=-i\int_{-\infty}^{\infty}\frac{dq}{2\pi}\frac{Z(q)}{q}\exp(iqz) \quad (10)$$

$V_{mn}(I,I')$ is then given by:

$$\begin{aligned}&V_{mn}(I,I')=-2\,\mathrm{Im}\int_0^{\infty}dq\frac{Z(q)}{q}G_m(q,I)G_n^*(q,I');\\&G_m(q,I)\equiv\int_0^{\pi}\frac{d\varphi}{\pi}\cos(m\varphi)\exp\left[iqz(I,\varphi)\right].\end{aligned} \quad (11)$$

Note that there is no bunch-to-bunch interaction in the formulas above - long-range wakes are omitted for the sake of simplicity. Equations (9-11) reduce the integro-differential BJV equation (4) to a standard eigen-system problem of linear algebra after the action integral in Eq.(9) is expressed as a proper sum.

## STEADY STATE SOLUTION

The algorithm of Eq. (6) allows the determination of a numerical solution of the steady state problem. In this section, the problem of the existence and uniqueness of that solution is considered.

It is well known that the steady-state solution does not necessarily exist. For example, below a certain temperature threshold, there is no thermodynamic equilibrium (no solution of the Haissinski equation [9]) for the space charge wake above transition, $W(z) \sim \delta(z)$ [8,10]. For this case though, the distribution function is given in the Hamiltonian domain, $F = C\exp(-H/T)$, not in the action domain, as in the previous section. As a consequence, the normalization constant $C$ for the Haissinski equation is yet to be found from the normalization condition: $2\pi C\int_0^{\infty}\exp(-H(I)/T)dI = 1$, which is to be added to the entire set of equations (5) and must be solved jointly with them. If the temperature $T$ is low enough, the normalization condition leads to an algebraic equation having no solutions. This situation is not specific to thermodynamic equilibrium only. A similar phenomenon appears for any distribution function in the Hamiltonian domain. For instance, the same problem emerges for the Hofmann-Pedersen distribution $F = C\sqrt{H_{\max}-H}$, as is shown in Ref. [8, Chapter 6.2]. For space charge above transition, and some other wakes, the bunch momentum spread and average Hamiltonian turn out to be limited from below: for a given RF and bunch population, they cannot be smaller than a certain value for any longitudinal emittance. That is why it may be wrong to assume an arbitrary distribution function in terms of the Hamiltonian. On the contrary, equations (5) assume distribution density as a function of action. Therefore it is a priori explicitly defined and normalized; thus, the specific limitations for Hamiltonian-domain distribution functions do not apply.

The following estimation shows when Eqs. (5) do have a solution. Let $\bar{I}$ be the rms bunch emittance, and $l$ be the rms bunch length. The rms momentum spread is then estimated to be $\bar{p} \cong \bar{I}/l$, and the average synchrotron frequency is $\Omega \cong \bar{p}/l \cong \bar{I}/l^2$. It is also known that $\Omega^2-\Omega_0^2 \propto \mathrm{Im}(Z(l^{-1}))/l^2$, where $\Omega_0$ is the bare RF synchrotron frequency (note that RF nonlinearity is neglected at this point). Combination of these two expressions yields (compare with Ref. [8], p. 285):

$$\bar{I}^2q^4 = 1+q^2\,\mathrm{Im}\,Z(q);$$

where $q=1/l$ is the inverse bunch length to be found from this equation; $q$ is measured in inverse radians of RF phase. The emittance $\bar{I}$ is dimensionless, and its value in conventional eV·s units can be found after multiplication by a factor of $E_0\Omega_0/(\eta\omega_{\mathrm{rf}}^2)$, where $E_0=\gamma mc^2$ is the beam energy. The dimensionless impedance $Z(q)$ of this paper, Eq. (10), relates to the conventional $Z_{||}(q)$ as $Z(q)=DZ_{||}(q)$ with the intensity factor $D = Nr_0\eta c\omega_{\mathrm{rf}}^2/\left(\Omega_0^2\gamma C\right)$, where $N$ is the bunch population, $r_0$ - the classical radius, $\eta=\gamma_t^{-2}-\gamma^{-2}$ – the slippage factor, $\omega_{\mathrm{rf}}$– RF angular frequency, $\gamma$ – relativistic factor, and $C$ – the machine circumference. Note that this equation does not give an exact solution for the bunch length. Instead, it is an estimate showing the existence of the solution and its dependence on the parameters. It follows that the solution exists if the wake is not too singular: at high frequencies the impedance may not grow too fast, providing $\lim_{q\to\infty}\mathrm{Im}\,Z(q)/q^2=0$, which is true for all realistic cases. For non-monotonic impedances there may, in general, be several solutions. For the space charge and the resistive wall impedances there is always a unique steady state.

However, since the RF potential is never an infinite parabola, Eqs. (5) may still have no solution. Indeed, the bucket has a limited acceptance; thus, it cannot hold a bunch with an emittance that is greater than that acceptance. Moreover, in many cases, wake fields reduce bucket capacity. This could lead to some beam loss to DC, even if the bunch fits within the bare RF bucket.

In a case where the distribution function is given as a function of Hamiltonian, with $\bar{H}$ as its average value, the steady state estimation for the parabolic RF is written as

$$\bar{H}q^2 = 1 + q^2 \operatorname{Im} Z(q).$$

In this case, the existence of a solution is not intensity-limited for slow-growing or bunch-lengthening impedances only, when $D \lim_{q\to\infty} \operatorname{Im} Z_{\|}(q) \le 0$; which is not satisfied for the space charge above transition and the resistive wall below transition. For these impedances, there are either no solutions, or there are two of them. In the case of two solutions, they have identical Hamiltonian distributions, but different phase space densities.

## VAN KAMPEN MODES

More than half a century ago, N. G. van Kampen found an eigensystem of the BJV equation for infinite plasma [1-3]. In general, this spectrum consists of continuous and discrete parts. The continuous spectrum essentially describes single-particle motion, accompanied with a proper plasma response. The frequency band of the continuous spectrum is one of the incoherent frequencies. For any velocity *v* within the distribution function, there is a continuous van Kampen mode with a frequency *kv,* where *k* is the wave number. Continuous modes are described by singular functions in the velocity space, underlying their primary relation to single-particle motion. In these terms, Landau damping results from phase mixing of van Kampen modes of the continuous spectrum. Unlike the continuous spectrum, discrete one not necessarily exists. If it does, the discrete modes are described by regular functions, and some of them do not decay with time. Indeed, since the original equations (analogue of Eq. (9)) possess real coefficients, the mode frequencies are either real or form complex-conjugate pairs. The first case corresponds to a pure loss of Landau damping (loss of LD, or LLD), while the second describes an instability. Plasma with a monotonic density distribution has been shown to be always stable [11]. The discrete modes of pure LLD type may only appear if the distribution function is of a finite width.

Most of the plasma results are applicable to circulating bunches in accelerators. However, two issues distinguish bunches and plasma. First, for beams, particle interaction may be described by various wake functions, being more diverse than pure Coulomb forces of the classical plasma. Second, the frequency spectrum for bunch particles is always limited, while in plasma the velocity spectrum may be considered infinite in extent, at least formally.

The eigenvalue problem of the BJV equation for bunch longitudinal motion was first considered by A. N. Lebedev [12]. Although the suggested formalism was not numerically tractable, an important result was analytically obtained. It was proved that for the space charge impedance above transition, a bunch steady state is always stable (which does not exclude LLD). The numerically tractable algorithm was suggested more than twenty years later by Oide and Yokoya [7].

For a parabolic RF potential, van Kampen modes were analyzed for power wakes [7], capacitive [13], broad-band wakes [7,13], and modified inductive wakes [14]. For that RF potential, rigid bunch oscillations at the unperturbed synchrotron frequency are always a solution of equation of motion [13]. Indeed, the single-particle equations of motion can be written as

$$\ddot{z}_i + \Omega_0^2 z_i = \sum_j W'(z_i - z_j)\,; \quad i, j = 1, \ldots, N\,.$$

The solution can be presented as a sum of a steady-state-related part $\hat{z}_i$ and a small perturbation $\tilde{z}_i$. It is clear that rigid-bunch mode, $\tilde{z}_i = \text{const} \cdot \cos(\Omega_0 t)$ satisfies this equation. While the rigid-bunch frequency is intensity-independent, all of the incoherent frequencies are typically either suppressed or elevated by the potential well distortion. Thus, this mode normally stays outside the incoherent band, and so is a discrete LLD-type mode. Although this is normally expected behavior, it is not necessarily exclusively the case. As was shown in Ref [13], for a broad band impedance model, core and tail incoherent frequencies may correlate with intensity in the opposite direction, so that the rigid-bunch mode may be covered by incoherent frequencies. Thus, the rigid-bunch mode should be Landau-damped in that case. In Ref. [15], loss of Landau damping was analyzed for the space charge impedance and various RF shapes, arbitrarily assuming the coherent motion as the rigid-bunch one. As is shown in the next section, that assumption is not correct when the RF frequency spread is taken into account. The action dependence of the emerging discrete mode is normally very different from that of the rigid-bunch case. Because of that, rigid bunch approximation overestimates the threshold intensity.

Without interaction, there are no discrete modes for Eqs.(9). All modes belong to a continuous spectrum, $\omega = m\Omega(I)$. If the bunch intensity is low enough, the weak head-tail approximation may be applied, allowing the omission of terms with different azimuthal numbers. In this case, it is straightforward to show that for monotonic distributions, $dF/dI<0$, and for symmetric potential wells, $U(-z)=U(z)$, Eq. (9) reduces to one with a symmetric matrix. In that case all of its eigenvalues are real. Since there are no unstable modes, all of the discrete modes, if any, belong to the pure LLD type. In practice there are always some energy losses, and so the distorted potential well $U(z)$ is always somewhat asymmetric. However, my attempts to find the weak head-tail instability in numerical solutions for monotonic distributions over frequencies, some realistic wakes and purposely asymmetric RF potentials have not yet succeeded. If the frequency distribution is not monotonic, a mode coupling instability is possible. To save CPU time, the stability analysis was limited by the weak head-tail approximation and considering only the dipole azimuthal mode. In other words, for the following analysis only the $m$=$n$=1 matrix elements are left in Eq.(9).

This paper takes into account two possible reasons for bunch intensity limitations: reduction of the bucket acceptance by wake fields and the emergence of a discrete mode (LLD).

## RESISTIVE WALL IMPEDANCE

In this section, the intensity limitations for resistive wall impedance are summarized. The beam energy is assumed to be above transition. The RF potential is written as

$$U_{\rm rf}(z) = (1-\cos z) + \alpha_2 (1-\cos 2z)/4 . \qquad (12)$$

Three options for the second RF harmonic are considered: single harmonic (SH) with $\alpha_2$=0, bunch shortening (BS) with $\alpha_2$=1, and bunch lengthening (BL) with $\alpha_2$=-1. Equation (12) assumes the synchrotron frequencies are given in units of zero-amplitude synchrotron frequency provided by the first RF harmonic only, $\Omega_0$. For the SH case, the RF bucket acceptance (maximal action) in these dimensionless units is $8/\pi \approx 2.54$. The energy offset is related to the dimensionless momentum by $\delta E / E_0 = -p\Omega_0 / (\eta\omega_{\rm rf})$. The time offset is $z/\omega_{\rm rf}$. The dimensionless wake function and impedance of a round chamber with radius $b$ are written as :

$$W(s) = -k/\sqrt{-s} \;;$$

$$Z(q) = k(1 - i\,{\rm sgn}\,q)\sqrt{\pi\,|q|/2} \;;$$

$$k = \frac{N r_0 \eta \omega_{\rm rf}^2}{\pi \gamma b \Omega_0^2} \sqrt{\frac{\omega_{\rm rf}}{\sigma}} \quad ,$$

where the wall conductivity $\sigma$ stays in the CGS units of 1/s.

An example with the parabolic RF potential suggests that wake fields act more on incoherent frequencies than on the coherent ones. For the parabolic potential, the first discrete mode does not depend on the impedance at all. Thus, at a certain threshold, a first discrete mode comes off the continuous spectrum, since its frequency is not suppressed or increased as much as the incoherent frequencies are. For the SH and BS RF, above transition, mostly lowest-amplitude particles are excited for this mode, since their frequencies are closer to the coherent frequency.

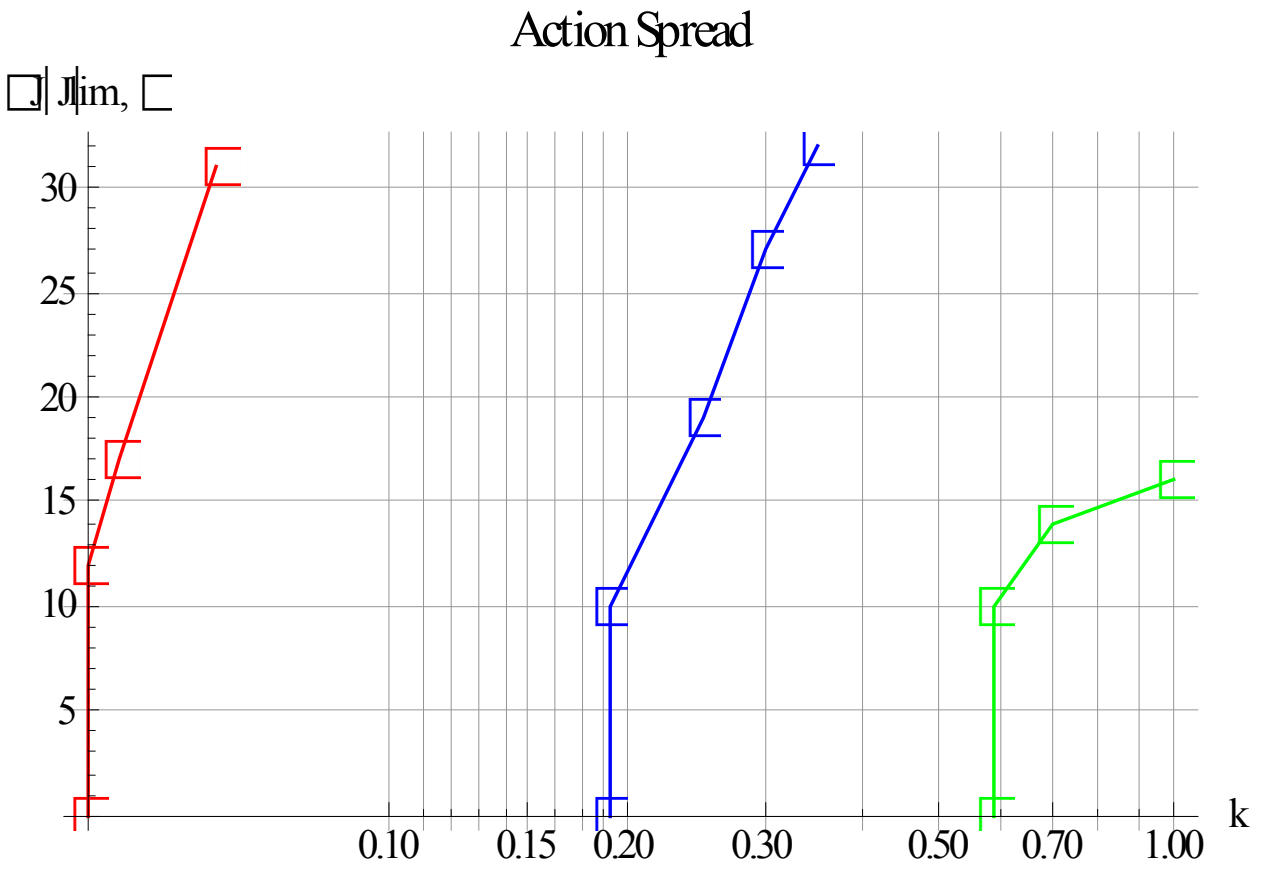

Figure 1: Relative width of the discrete mode $\sigma_I/I_{\rm lim,\ \%}$, versus the intensity parameter $k$ for the distribution $F(I)$~$(I_{\rm lim}$-$I)^{1/2}$ and emittances $I_{\rm lim}$=0.5 , 1.0 and 1.5 (red, blue and green), SH case.

An assumption of rigid-bunch discrete mode normally is far from being correct. In Fig.1, the relative width of that discrete mode $\sigma_I/I_{\rm lim}$ is shown as a function of the intensity parameter $k$ for 3 emittances and the Hoffman-Pedersen distribution $F(I)$~$(I_{\rm lim}$-$I)^{1/2}$. The relative width is defined as

$$\sigma_I = \sqrt{\frac{\int dI f^2(I)(I-\bar{I})^2}{\int dI f^2(I)}}, \; \bar{I} \equiv \frac{\int dI f^2(I) I}{\int dI f^2(I)} .$$

The mode widens rather fast above threshold as a consequence of being singular at the threshold. All of the modes are primarily located at small amplitudes, $\bar{I} \ll \sigma_I$.

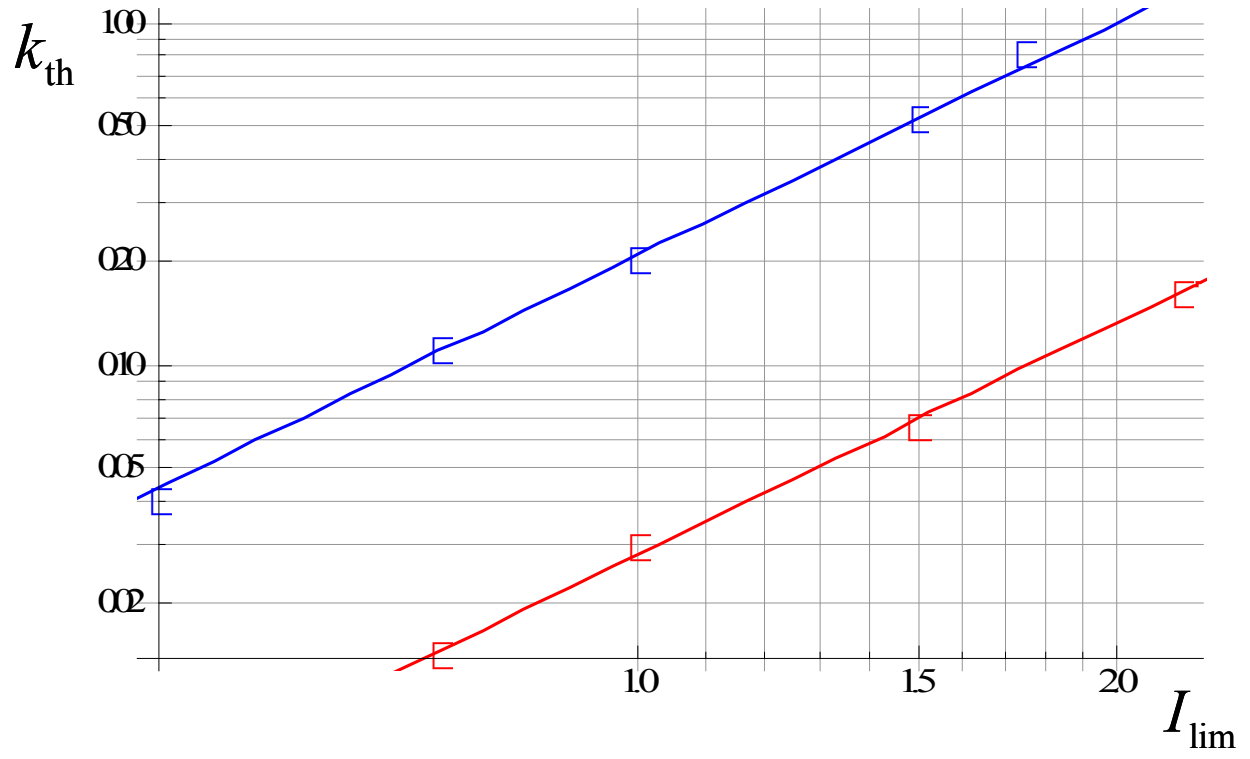

Figure 2: LLD threshold of the resistive wall intensity parameter $k_{\rm th}$ for SH RF, versus the bunch emittance for $F(I) \propto (I_{\rm lim} - I)^P$ Blue dots : $P$=1/2, red dots: $P$=2, lines are fits with $k_{\rm th} \propto I_{\rm lim}^{9/4}$. Note the strong dependence on the distribution parameter $P$.

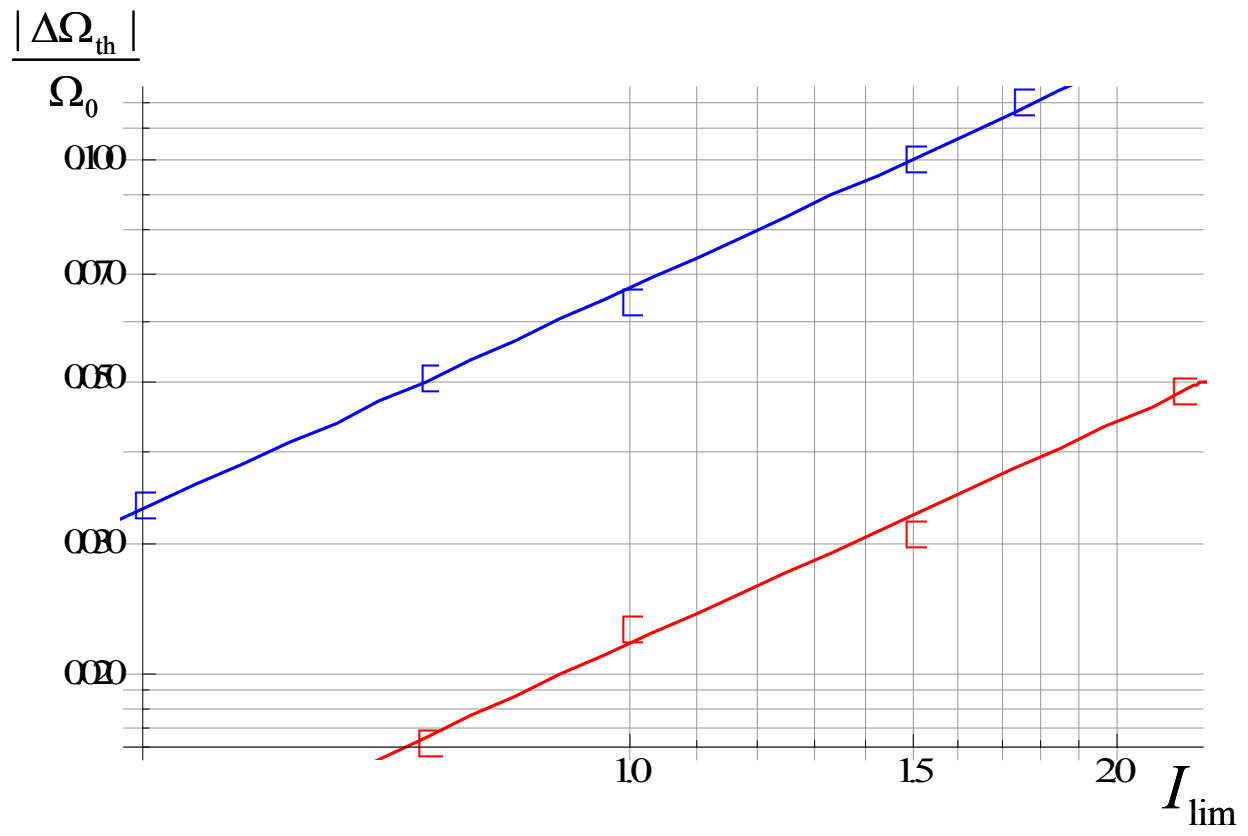

Figure 3: Same thresholds as in Fig. 2, in terms of the relative shift of zero-amplitude synchrotron frequencies $\Delta\Omega = \Omega(0) - \Omega_0$. Lines are linear fits $\Delta\Omega_{th} \propto I_{lim}$. Note that the threshold frequency shifts are fairly low, even for the blue line.

LLD thresholds for the intensity parameter $k$ versus emittance $I_{lim}$ for two different distributions are presented in Fig.2. The power law $k_{th} \propto I_{lim}^{9/4}$ agrees with a dependence obtained by method of Ref. [17], i. e. a comparison of the zero-amplitude synchrotron tune shift $\Delta\Omega = \Omega(0) - \Omega_0 \propto \mathrm{Im}\, Z(l^{-1}) / l^2$ with the synchrotron tune spread in the nonlinear SH RF, $\delta\Omega \propto l^2$. Equating these two values with the bunch length $l \propto \sqrt{I_{lim}}$, and $\mathrm{Im}\, Z(l^{-1}) \propto l^{-1/2}$ one gets the thresholds $k_{th} \propto I_{lim}^{9/4}$ for the resistive wall impedance. For inductive impedance, it gives $k_{th} \propto I_{lim}^{5/2}$, as was found in Ref. [17]. For any impedance, at the LLD threshold $\Delta\Omega \cong \delta\Omega \propto I_{lim}$. Although this scaling is based on the small bunch frequency spread formula, $\delta\Omega \propto l^2$, it appears to be valid up to full bucket case, see Figs. (2, 3). A reason for that is that the discrete mode in this case appears above all the incoherent spectrum, so it is mostly associated with the low-amplitude particles. That is why the mode is mostly sensitive to the frequency spread of those particles, for which the small-bunch approximation $\delta\Omega \propto l^2$ is always valid.

Contrary to the SH and BS RF cases, for the BL case incoherent frequency is not a monotonic function of action, it has a maximum at $I = I_m \approx 1.5$. That is why, for the considered case of an effectively repulsive wake, the discrete mode emerges from the tail particle frequencies if the bunch limiting emittance is small, ($I_{lim} < I_m$). For BL RF, the emergence of the discrete mode is sensitive to the tails of the distribution: even a tiny tail covering the coherent frequency yields Landau damping, killing that discrete mode. If the bunch emittance is not that small ($I > I_m$), the discrete mode emerges above the incoherent maximum. Since this mode emerges outside the entire bucket area of the incoherent frequencies, Landau damping cannot be restored by tiny perturbations of the distribution function. That is why this kind of LLD, which cannot be cured by tiny corrections of the distribution function, is called hereinafter "radical LLD." To avoid that tail ambiguity, only radical LLD is taken as a real stability limit here.

Figure 4 shows radical LLD limitations for BL RF, for two different distributions. Note that LLD restricts the available acceptance by $I_m \approx 1.5$, while the entire BL bucket area is about twice higher. Stability limitation associated with the maximum of incoherent synchrotron frequencies was first pointed out in Ref. [19]; it was studied at CERN SPS (see Ref. [20] and references therein).

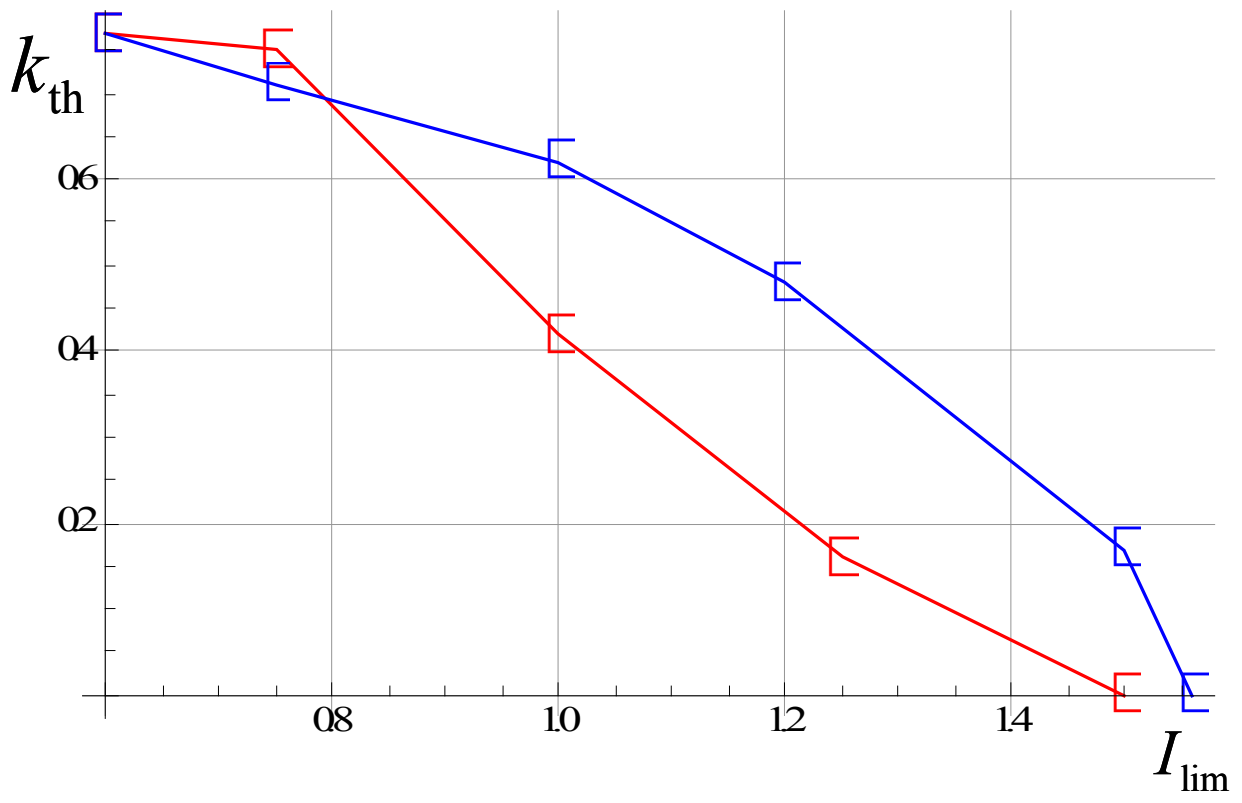

Figure 4: Threshold intensity $k_{th}$ versus emittance $I_{lim}$ for BL RF and two distribution functions: $F(I) \propto (I_{lim} - I)^{1/2}$ (red) and $F(I) \propto (I_{lim} - I)^2$ (blue).

On the $k$-$I_{lim}$ area, the availability is limited by LLD and the bucket acceptance. For the three RF configurations, SH, BS and BL, their areas of availability are shown in Fig. 5 with $F(I) \propto (I_{lim} - I)^{1/2}$. Clearly, every RF configuration has its own beneficial area: hot and low-intensity beams better fit into SH, cold high-intensity ones are more suitable for BL, and the intermediate case is for BS RF.

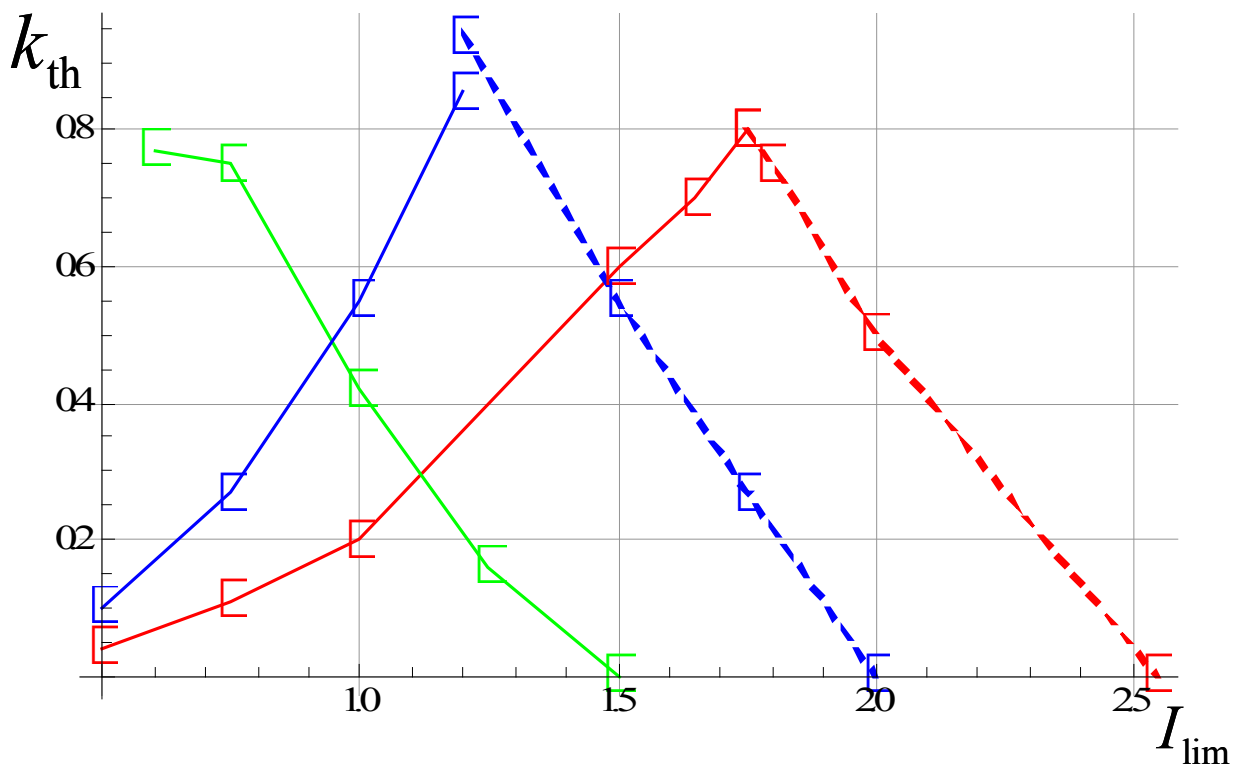

Figure 5: Intensity-emittance $k$-$I_{lim}$ areas of availability for $F(I) \propto (I_{lim} - I)^{1/2}$. Red lines are for SH, blue – for BS, green – for BL. Solid lines show radical LLD, dashed – limiting bucket capacity.

## INDUCTIVE IMPEDANCE

Hadron machines are normally dominated by resistive wall and inductive (or space charge) impedances (see e. g. Ref. [19]). In the dimensionless units, the inductive wake function and impedance are written as

$$W(s) = -k\delta(s)\,;$$

$$Z(q) = -ikq\,;$$

$$k = -\frac{2Nr_0\eta\omega_{\rm rf}^3}{\gamma c\Omega_0^2}\frac{\operatorname{Im} Z_{\|}}{nZ_0}\quad ,$$

where $Z_0$=4$\pi$/c=377 Ohm, and $n$ is the revolution harmonic number. LLD threshold lines, $k_{\rm th}$ versus $I_{\rm lim}$ , are presented in Fig (6) for SH RF, $k$>0 and three distribution functions: $F(I)\propto(I_{\rm lim}-I)^{1/2}$ (most stable), $F(I)\propto(I_{\rm lim}-I)^2$ (medium stable) and $F(I)\propto(I_{\rm lim}-I)^2\left(1+\cos(8\pi I/I_{\rm lim})\right)$ (least stable). The last distribution mimics a coalesced proton bunch in the Tevatron. It takes about an hour for memory of the constituent 7 bunches be smeared in the coalesced proton bunch in the Tevatron.

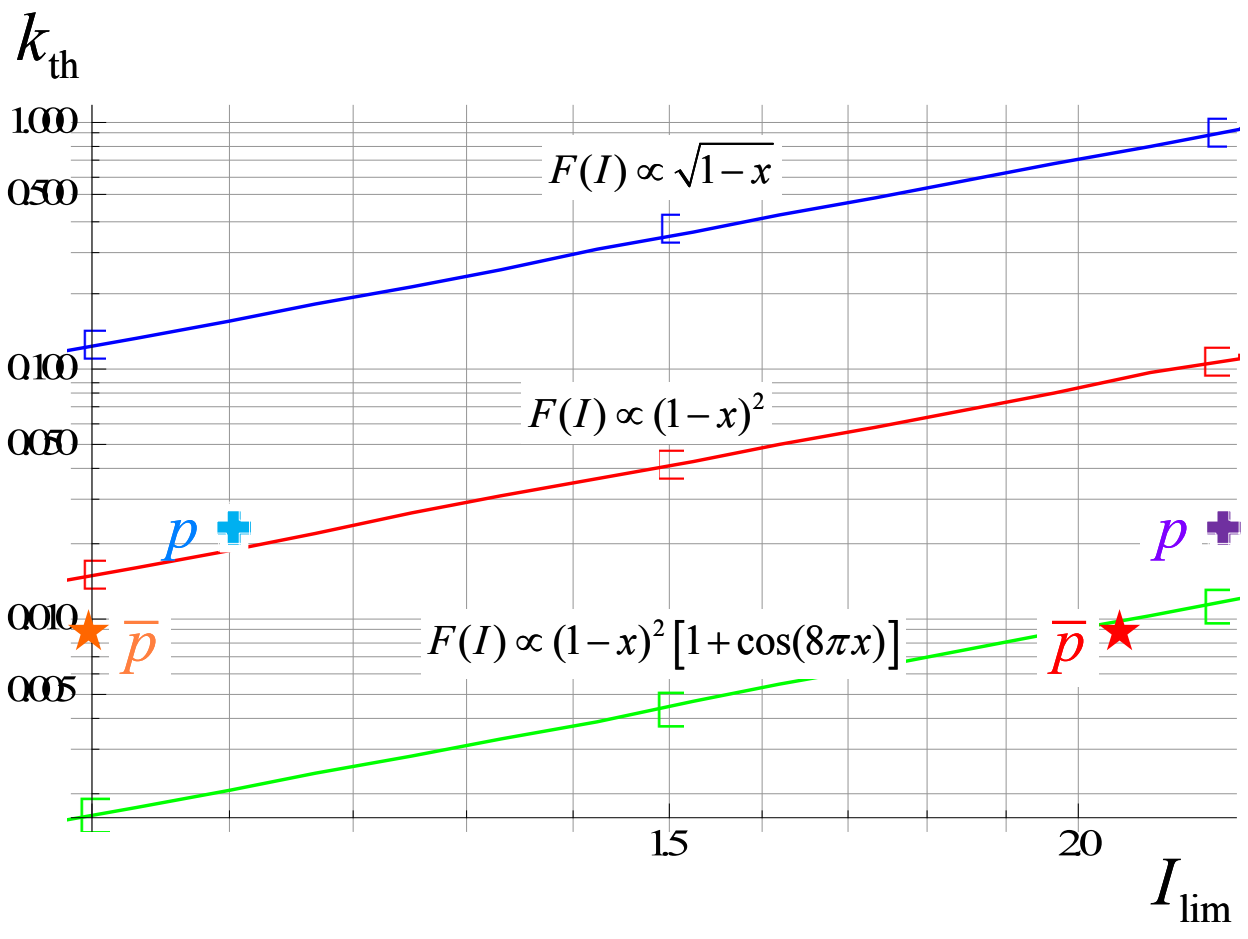

Figure 6: Threshold LLD intensity parameter $k_{\rm th}$ for SH RF versus the bunch emittance for 3 denoted distributions $F(I)$, where $x = I/I_{\rm lim}$. Lines are fits with $k_{\rm th}\propto I_{\rm lim}^{5/2}$. Crosses and stars show protons and antiprotons at Tevatron at injection (violet and red, on the right part) and top energy (blue and orange, on the left).

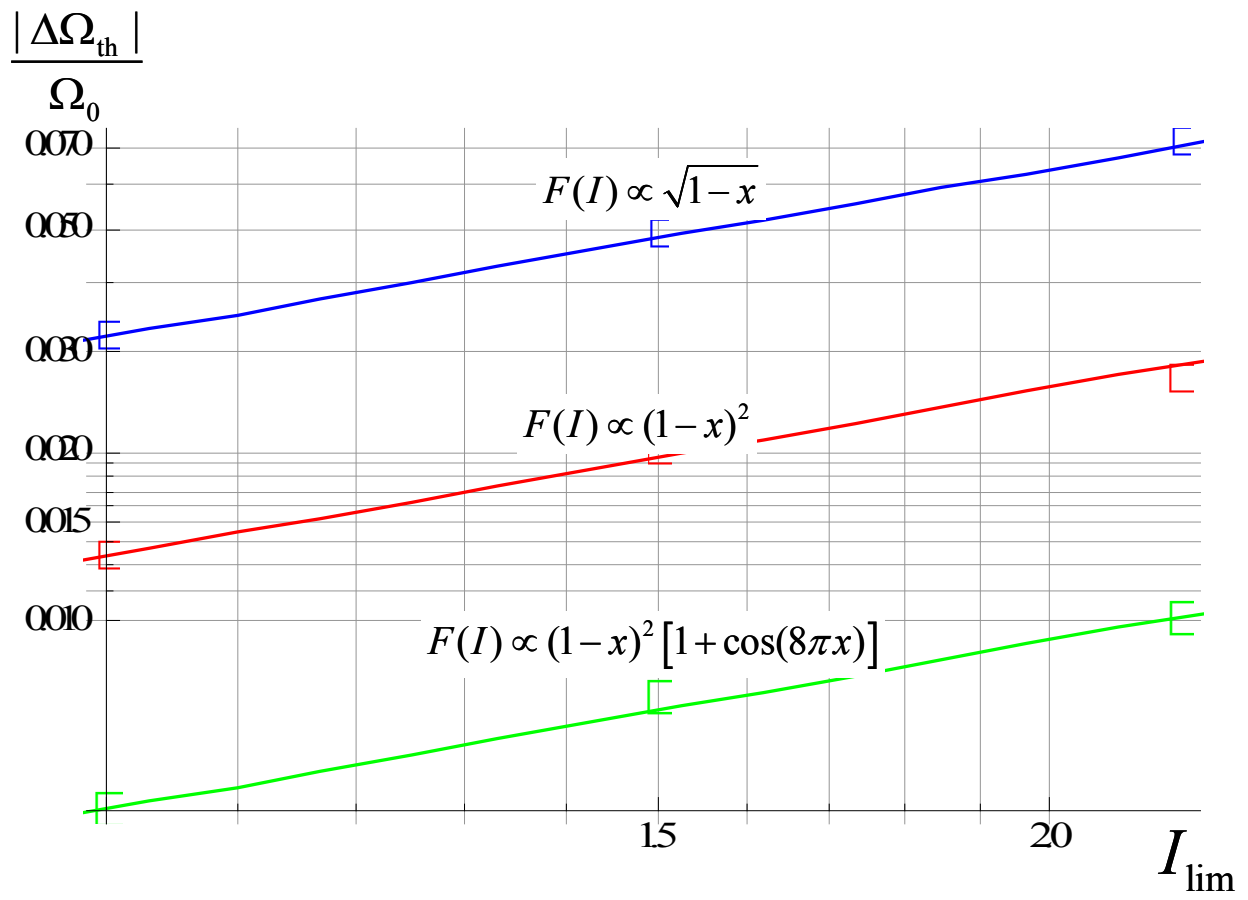

Figure 7: Same thresholds as above, in terms of the zero-amplitude relative incoherent frequency shifts. Lines are linear fits.

Threshold dependences $k_{\rm th}\propto I_{\rm lim}^{5/2}$, $\Delta\Omega_{\rm th}\propto I_{\rm lim}$ are in agreement with scaling low of Ref. [17]. The plots in Figs. (2,3,6,7) show the strong dependence of the threshold intensity on details of the distribution function. Qualitatively, this can be interpreted as a high sensitivity to steepness of the bunch distribution at small arguments. That high sensitivity should not be too surprising. Indeed, for both resistive wall and inductive impedances, the threshold phase space density $k_{\rm th}/I_{\rm lim}$ goes down with the bunch length. It may sound counter-intuitive, but it is what is shown above: at a given phase space density, the least stable are the less populated bunches! A reason for that follows from a fact that, at fixed phase space density, the less populated bunch is shorter. The wake effect is stronger for shorter bunches, and the stabilizing frequency spread is weaker for them. That is why a small central portion of a bunch is less stable than the entire bunch. The ability of that small portion to be effectively disconnected from the entire bunch depends on the distribution function: the steeper the distribution at small amplitudes, the stronger this ability is.

According to the Tevatron impedance model [21], its proton bunches at injection are 2 times above the green line LLD threshold of Fig. (6). At the top energy, they are ~20 times above that threshold. This agrees with the observations: in the reality, the protons are always "dancing" in the Tevatron, unless the damper is on [18]. According to these calculations, the antiprotons stay slightly below the green line threshold at injection, and they are ~10 times above it at the top energy. In reality, they are stable at injection, and unstable at collisions. To conclude, both proton and antiproton stability observations are in agreement with the model described.

Since the LLD threshold strongly depends on the small-argument steepness of the distribution function, its local flattening there should increase the LLD threshold. That local flattening can be realized by means of RF phase modulation at a narrow frequency band around the zero-amplitude synchrotron frequency, smearing the distribution for the resonant particles. Dedicated experiments with this bunch shaking were performed at the Tevatron; it was realized that this bucket shaking indeed stops bunch dancing, see Ref. [22].

It may be supposed that LLD is a reason for the persistent bunch oscillations observed at RHIC [23, 24] and CERN SPS [20] as well. To the author's point of view, the LLD explanation is rather a supplement than a competitor to the soliton theory of Refs. [23, 24]. Indeed, being linear, LLD describes a mode growth from whatever small initial perturbation due to multi-turn or couple-bunch wakes. This growth can be saturated at nonlinear stage of the coherent motion, where the soliton theory takes over.

## CONCLUSIONS

This paper includes a general treatment of a bunch self-consistent steady state in a distorted potential well, and van Kampen mode analysis for the steady state case. Criteria of existence and uniqueness for the steady state problem in action and Hamiltonian domains are formulated. A relaxation method for numerical solution of the steady state problem is described and used.

The language of van Kampen modes is a powerful tool for studying beam stability. This method leads to an eigensystem problem, so it is straightforward to implement numerically. Sum of the growth rates of the emergent discrete modes is zero; thus, some of them do not decay; there is loss of Landau damping for them. By definition, the discrete modes lie outside the continuous incoherent spectrum, but they may still stay within the bucket. In the last case, the discrete mode would disappear after a tiny portion of resonant particles are added. However, if the discrete mode lies outside the bucket, Landau damping cannot be restored by a tiny perturbation of the particle distribution; this sort of LLD is characterised as radical..

For a given bunch emittance and RF voltage, the intensity is limited either by reduction of the bucket acceptance or by (radical) LLD. In this paper, results are presented for longitudinal bunch stability in the weak head-tail approximation for resistive wall and inductive impedances. For the resistive wall impedance, three RF configurations are studied: single harmonic (SH), bunch shortening (BS) and bunch lengthening (BL). It is shown that every one of these RF configurations may be most preferable, depending on the bunch emittance and intensity.

The LLD threshold intensities are typically very low. For the cases under study, the threshold low-amplitude incoherent frequency shifts vary from 10% to 1% at full bucket. Although LLD itself means in many cases the emergence of a mode with zero growth rate, even tiny multy-turn or couple-bunch wake can drive the instability for the discrete mode.. In that sense, LLD is similar to the loss of the immune system of a living cell, when any microbe becomes fatal for it.

Specific results of this paper agree with the power lows for LLD suggested by Ref.[17]. However, the numerical factors obtained here for these lows strongly depend on the bunch distribution function. Particularly, for SH RF and inductive impedance above transition, for the three examined distributions, the highest LLD threshold intensity exceeds the lowest one by almost two orders of magnitude.. Based on that observation, a method of beam stabilization is suggested [22].

The author is thankful to Valeri Lebedev (FNAL) and Slava Danilov (ORNL) for multiple productive discussions.

## REFERENCES

[1] N. G. van Kampen, Physica (Utrecht) 21 (1955) 949.
[2] N. G. van Kampen, Physica (Utrecht) 23 (1957) 647.
[3] G. Ecker, “Theory of Fully Ionized Plasmas”, Academic Press, 1972.
[4] A. Burov, “Van Kampen modes for bunch longitudinal motion”, Proc. HB 2010, Switzerland.
[5] A. Burov, “Dancing bunches as van Kampen modes”, Proc. PAC'11, New York.
[6] M. Heńon, Astron. Astrophys., 114, 211-212 (1982).
[7] K. Oide and K. Yokoya “Longitudinal Single-Bunch Instability in Electron Storage Rings”, KEK Preprint 90-10 (1990).
[8] A. Chao, “Physics of Collective Beam Instabilities”, Whiley Interscience, 1993.
[9] J. Haissinski, Nuovo Cimento 18B, No. 1, 72 (1973).
[10] A. Burov, “Bunch lengthening, self-focusing and splitting”, Thesis (in Russian) (1991).
[11] O. Penrose, Phys. Fluids, **3**, 258 (1960).
[12] A. N. Lebedev, Atomnaya Energiya 25 (2), p. 100 (1968).
[13] M. D'yachkov and R. Baartman, Part. Acc. 50, p.105 (1995); also TRIUMF preprint TRI-PP-94-45 (1994).
[14] Y. Shobuda and K. Hirata, Phys. Rev. E, v. 60 (2), p. 2414 (1999).
[15] O. Boine-Frankenheim and T. Shukla, Phys. Rev. ST-AB, 8, 034201 (2005).
[16] A. Chao, B. Chen and K. Oide, “A weak microwave instability with potential well distortion and radial mode coupling”, Proc. PAC 95, p. 3040 (1995).
[17] A. Hofmann and F. Pedersen, IEEE Trans. Nucl. Sci. **26**, 3526 (1979).
[18] R. Moore et al., “Longitudinal bunch dynamics in the Tevatron”, Proc. PAC 2003, p. 1751 (2003).
[19] V. I. Balbekov and S. V. Ivanov, “Longitudinal beam instabilities in proton synchrotrons”, Proc. 13th Int. Conf. On High-Energy Acc., Novosibirsk, 1987, v. 2, p.127 (in Russian).
[20] E. Shaposhnikova, “Cures for beam instabilities in CERN SPS”, Proc. HB 2006 Conf., Tsukuba, 2006.
[21] Fermilab report, “Run II Handbook” http://www-bd.fnal.gov/lug/runII_handbook/RunII_index.html , p. 6.58 (1998).
[22] A. Burov and C-Y Tan, “Bucket shaking stops bunch dancing in Tevatron”, Proc. PAC'11, also submitted to Phys. Rev. ST Accel. Beams.
[23] M. Blaskiewicz et al.,“Longitudinal solitons in RHIC”, Proc. PAC 2003, p. 3029.

[24] M. Blaskiewicz et al., Phys. Rev. ST Accel. Beams **7**, 044402 (2004).